\documentclass[usenatbib]{mnras}
\usepackage[T1]{fontenc}
\usepackage{amsmath}
\usepackage{multirow}
\usepackage{graphicx}

\begin{document}
\title{Cyclic variations of the structure and energetics of solar magnetic fields}
\author[V.~Obridko et al.]{V.N.~Obridko$^{1,2}$\thanks{obridko@izmiran.ru}, A.S.~Shibalova$^{1}$, D.D.~Sokoloff$^{1,3,4}$\thanks{sokoloff.dd@gmail.com}, \\
$^{1}$IZMIRAN, 4, Kaluzhskoe  Shosse, Troitsk, Moscow, 108840, Russia\\
$^{2}$Central Astronomical Observatory of the Russian Academy of Sciences at Pulkovo, St.Petersburg, Russia\\
$^{3}$Department of Physics, Lomonosov Moscow State University, Moscow, 119991, Russia\\
$^{4}$Moscow Center of Fundamental and Applied Mathematics, Moscow,
119991, Russia}

\date{Received ..... ; accepted .....}
\maketitle

\begin{abstract}
The solar cycle is a complex phenomenon, a comprehensive understanding of which requires the study of various tracers. Here, we consider the solar cycle as manifested in the harmonics of the solar large-scale surface magnetic field, including zonal, sectorial and tesseral harmonics, divided into odd and even relative to the solar equator. In addition to considering the amplitudes of the harmonics, we analyze their contribution to the magnetic energy. It turns out that the relative contribution of different types of harmonics to the magnetic energy is virtually independent of the cycle height. We identify different phases of the activity cycle using harmonics of different symmetries. A possible way to incorporate the obtained result into the solar dynamo theory is proposed.
\end{abstract}


\section{Introduction}

According to the generally accepted model of magnetic field generation on the Sun, under the influence of differential rotation, the original poloidal field is transformed into a toroidal one. Then the meridional circulation or movement of the tail magnetic field towards the poles restores the poloidal nature of the global magnetic field. Simultaneously with the moment of maximum of the poloidal field, local fields pass through a minimum. At this moment, the sign of the local fields changes (Hale’s law), and at the moment of maximum local fields, a reversal of the field at the pole occurs. This means that the real physical period of the solar activity cycle is not 11, but 22 years.

This work is devoted to the analysis of the structure and evolution of the large-scale magnetic field on the Sun. In this case, the large-scale field itself consists of multipoles of different orders. These multipoles are independent to a certain extent, and our task is to trace their evolution and interaction. The amplitudes and phases of the lower-order multipoles and their correlation with the solar photospheric magnetic field have been investigated in many studies (e.g., \cite{L77, H84, SF86, SW87, H91, GJ92, Getal92, S94, KS05}) on the basis of Kitt Peak and WSO (John Wilcox Observatory, Stanford) data.

In recent years, the Babcock-Leighton dynamo model has become most widespread. It really fits the observations best (e.g., \cite{L20}). The polar magnetic field left over from the previous cycle changes its sign during the sunspot growth phase \citep{Wetal89, B04, DEetal10}.

It turned out that polarity reversals in the Babcock–Leighton 
model (that is, the coupling of modes of different parities) well describe the observed polarity changes at the equator and at the poles 
of the Sun (\cite{DRetal12}).  In this case, the poloidal magnetic field can arise either as a result of the evolution of inclined bipolar regions, or as a result of generation from the toroidal field due to the {alpha}-effect throughout the entire convective zone. The latter source is a common 
component in distributed mean-field dynamo models (see results obtained by \cite{Petal14, Hetal14}, as well as the updated version of the Babcock–Leighton dynamo model presented by \cite{CS17}.  Note that the distributed dynamo paradigm is supported by a number of the global convection simulations (see, e.g.. \cite{Ketal16, Wetal18}). A comparison of the distributed dynamo model and the Babcock–Leighton scenario is given in \cite{B05, Ketal13}.

To summarize, we see that the theory of generation of a magnetic cycle has successfully addressed mainly the basic model of a cycle in relative units, while the dynamo as the main process of the extended cycle still remains not entirely clear (cf. \cite{Oetal23}). In particular, it is still not clear why the cycles differ in amplitude. A certain progress has been made in predicting the cycle amplitude on a time scale of several years based on some precursors, including the polar magnetic field. (See \cite{OS09, OS17} for early publications on the topic and \cite{N21, Netal21} for  comprehensive reviews. The methodology of prediction of the amplitude of the forthcoming cycle  based on the polar field data is described in detail in \cite{P20}). The method seems physically reasonable, still its statistical validity is based on the experience of only four cycles.

 On the other hand, the point concerning the self-similarity of the cycle structure (i.e., whether the structure is determined by the cycle amplitude alone) requires clarification. Note that the mean-field dynamo models take into account only the structures of the large-scale magnetic field,  while the development of sunspots and active regions from the large-scale magnetic field is still under investigation. The basic mean-field dynamo models do not include the sunspot-scale features. It is believed that the mean-field dynamo creates a toroidal magnetic flux somewhere near the solar surface; however, the flux transport in the sunspot magnetic field still needs clarification. Note that an interesting idea may be the negative magnetic pressure \citep{Ketal89, Betal10} or the thermal instability \citep{KM00}.

Solar activity since its discovery has generally been understood as the evolution of local magnetic fields. It is easier to observe and can be characterized by several well-known indices. This is the number and area of sunspots, the number of flares and radio bursts, the characteristics of active regions, torches and flocculi, as well as integrated radio, X-ray and UV fluxes and many others. At the same time, there is a clear shortage of indices characterizing cyclical changes in the structure and energy of global magnetic fields.

Below, we are using a spherical analysis as a tool to determine the contribution of various structural features to the cyclic large-scale evolution of the magnetic field. The analysis is based on magnetic field data obtained at Stanford with 3 arc min resolution for the period from June 1976 to July 2022, i.e., for Carrington rotations Nos. 1642—2055 (http://wso.stanford.edu/forms/prsyn.html). 
 
Instead of using the directly measured data, we represent them in the form of the Legendre polynomials for the convenience of calculations. It is assumed that, in the spherical layer from the surface of the photosphere to a certain spherical surface conventionally called the source surface, there are no currents. Then, the structure is completely described by the potential approximation as follows:

\begin{eqnarray}
 B_r = \sum_{l,m} P_l^m (\cos \theta) (g_l^m \cos m\phi + h_l^m \sin m \phi ) ((l+1) \times \cr
 \times (R_0/R)^{l+1} - l ((R/R_s)^{l+1} c_l),
 \label{br}
 \end{eqnarray}
 \begin{eqnarray}
 B_\theta = - \sum_{l,m} \frac{\partial P_l^m (\cos \theta ) } {\partial \theta} (g_l^m \cos m \phi + h_l^m \sin m \phi) \times \cr
\times ((R_0 /R)^{l+2} + (R/R_s)^{l-1}c_l), 
\label{btheta}
\end{eqnarray}
\begin{eqnarray}
B_\phi = - \sum_{l,m} \frac{m}{\sin \theta} P_l^m (\cos \theta) (h_l^m \cos m \phi - g_l^m \sin m \phi) \times \cr
\times ((R_0/R)^{l+2} + (R/R_s)^{l-1}c_l).
\end{eqnarray}
Here, $0 \le m$, $l < N$ (usually, $N \le 9$),
$c_l=-(R_0/R_s)^{l+2}$,  $P_m^l$  are the Legendre
polynomials, and $g_l^m$, $h_l^m$  are the harmonic coefficients. The latter was calculated from WSO Stanford data. To find the harmonic coefficients, $g_l^m$ and $h_l^m$, and thus, to fully determine the solution, we had to use the following boundary conditions. As the boundary conditions for the solution in the above form, it is necessary to know the radial component of the magnetic field at the lower boundary, i.e., at the photosphere level. The upper boundary is the source surface, where all field lines are radial. 
 
Traditionally, the evolution of the large-scale magnetic field is presented in the form of synoptic maps with a timescale of one Carrington rotation (see, e.g., Fig.~2 in \cite{Oetal23}). In this case, any dependence on longitude is averaged, and only the latitude-time dependence remains. Such a diagram is represented by zonal harmonics with $m=0$ only. In Fig.~1, which corresponds to Fig.~3 from \cite{Oetal23}, we show this diagram separately for the lower harmonics with $l=1, 3, 5$ (Fig.~1, upper panel) and the higher harmonics with $l=5, 7, 9$ (Fig. 1, lower panel). Note that the harmonic with $l=5$ is included in both plots, since it is considered a boundary separating both cases. 

\begin{figure}
\includegraphics[width=0.97\columnwidth]{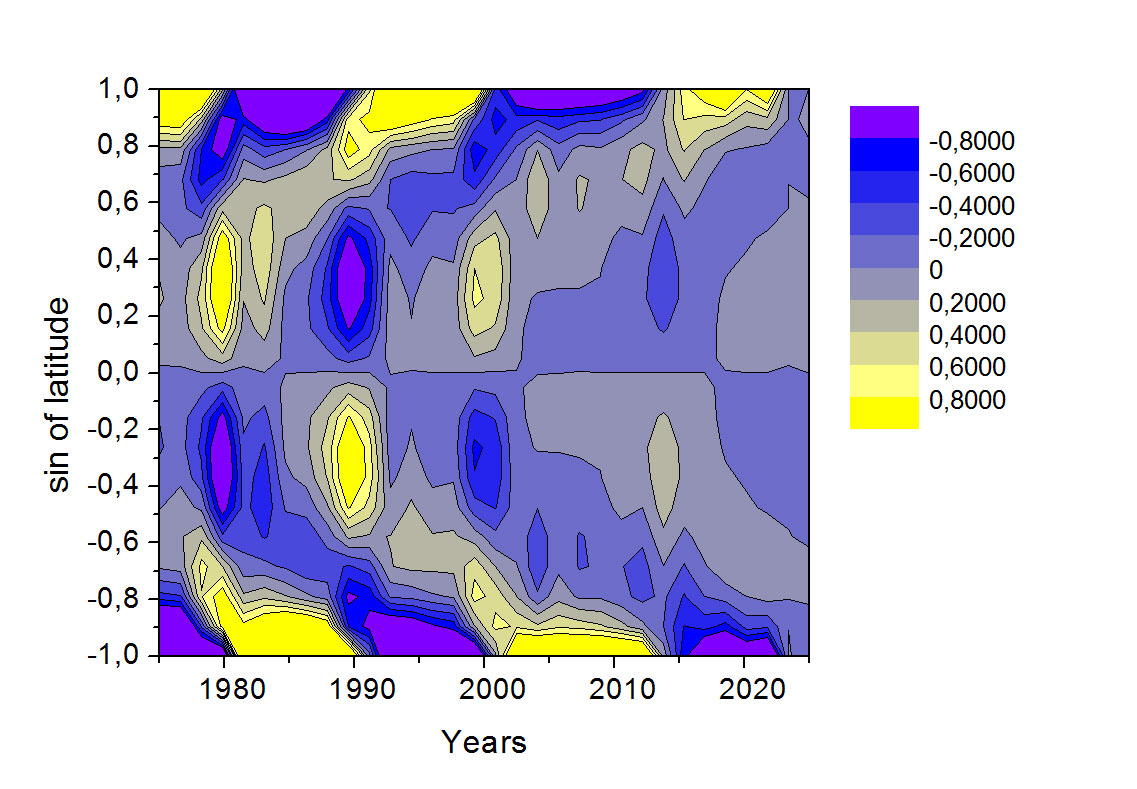}
a
\includegraphics[width=0.97\columnwidth]{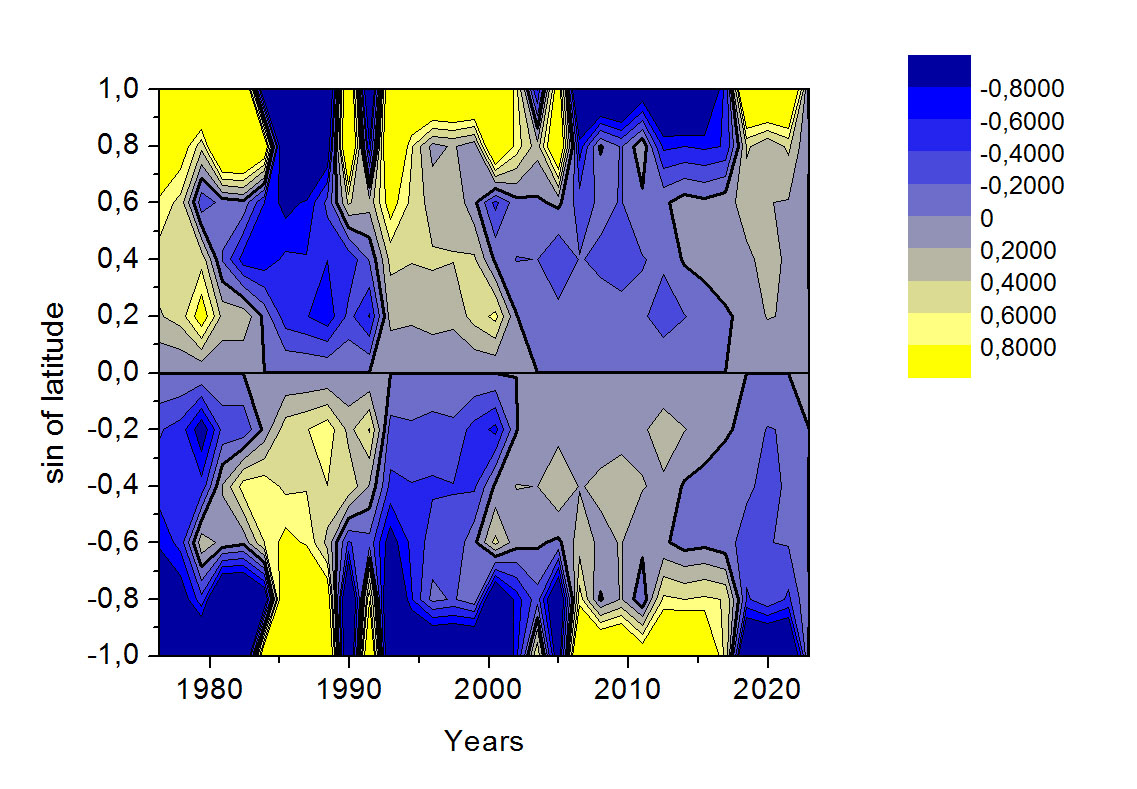}
b
\caption{The upper panel shows the time dependence of squared magnetic field connected with the lower zonal harmonics with 
$l=1, 3, 5$ 
while the lower panel shows the corresponding contribution of higher zonal harmonics  with
$l =5, 7, 9$.} 
\label{f1}
\end{figure}

Recently \cite{Oetal23} have proposed a further development of the traditional concept of the extended solar cycle. The basic solar dynamo model relies on the description of the 11-year solar activity cycle in terms of the first odd zonal harmonics of the large-scale solar magnetic field.  The point, however, is that this description itself needs some modification and development.  Indeed, during the transition from cycle to cycle, when the successive  activity wings on butterfly diagram overlap in time, three waves of activity coexist on the solar surface, leading to the formation and amplification of the odd harmonic with $l=5$. 

In order to explain the fact that one zonal harmonic is so specific to separate two types of behaviour of the magnetic features let us consider the synoptic map in the vicinity of 1999 (Fig.~1). Near the North pole, we can still see the wave of positive polarity, whose propagation began in 1989 (the growth phase of Cycle 22) and will continue for 13 years till 2002 (i.e., until the maximum of 23 cycle). Simultaneously, another wave of negative polarity started at mid latitudes in 1999 and will propagate to the pole until 2012 (i.e., until the growth phase of Cycle 24). 
The third wave starts in 1999. Unlike the previous two waves, it is not strictly unipolar; however, the positive polarity dominates there corresponding to the polarity of the leading spots in active regions. This wave propagates to the equator.  As a result, three activity waves with their specific polarities are present at a time. The coexistence and interaction of three magnetic configurations lasts from 1989 to 2005, i.e. covers 16 years. So, we have six zones on the whole visible solar disc with the polarities changing from wave to wave. This description exactly corresponds to the zonal harmonic with $l=5$. We call this time interval "the overlapping phase".   

\cite{Oetal23} presented the phenomenology of the extended cycle and \cite{Ketal21} connected the phenomenon with physical processes in the convection zone basing on helioseismological data for the last two activity cycles. It is assumed that the surface phenomena mentioned above are associated with the migration of the magnetic field from the bottom of the convection zone to the solar surface. Within this model, the phenomenon of the extended solar cycle is due to the quenching of the convective heat flux by the magnetic field and the modulation of the meridional circulation  by the heat flux variations. In addition, the subsurface rotational shear layer (leptocline, {\bf see e.g. \cite{GR01}}) plays a key role in the formation of the magnetic butterfly diagram.

We see that different harmonics of the surface solar magnetic field bear significant information about the solar activity cycle. However, the above consideration is still focused on the zonal harmonics. We think it's worth considering the entire set of harmonics in search for what additional information they can provide concerning the structure and origin of the solar activity cycle. This is the aim of our paper.

Obviously, we can calculate the index $i(B_r)$ for the contribution to the magnetic energy, for example, separately from the zonal, sectorial, and tesseral harmonics. The quantities are denoted by the corresponding subscripts.

\section{Different contributions to the total magnetic field energy}

To analyze cyclic variations in the global magnetic field of the Sun, it is convenient to introduce magnetic indices, which are quantities determined by the global distribution of the magnetic field.

Let $i(B_r)$ be the square of the radial component of the magnetic field averaged over the solar surface:

\begin{equation} 
i(B_r)|_R = <B_r^2>,
\label{ind}
\end{equation}
where $R$ is the radius of the sphere over which the averaging is performed.

In order to calculate $i(B_r)$, we have to know the magnetic field $B_r$, Eq.~(\ref{br}). After that, 
we can calculate the field at any point of a spherical layer. Applying the averaging procedure (\ref{ind}) to Eq.~(\ref{br}),  we obtain expressions for $i(B_r)$, 
which are essentially the mean square of the radial magnetic field on any given spherical surface from the photosphere to the source surface  \citep{OY89, OS92}. 
In particular,  we obtain the following expressions  for the  photosphere and the source surface,

\begin{equation}
i(B_r)|_{R_\odot} = \sum_{l,m} \frac{(l+1+l \zeta^{2l+1})^2}{2l+1}((g_l^m)^2 + (h_l^m)^2),    
\label{iph}
\end{equation}
\begin{equation}
i(B_r)|_{R_s} = \sum_{l,m} (2l+1) \zeta^{2l+4} ((g_l^m)^2 + (h_l^m)^2),
\label{is}
\end{equation}
where $\zeta = R_0/R_s$. This means that the contribution of the $l$-th mode to the mean magnetic field contains an $l$-dependent coefficient. In these formulas, $i(B_r)|_{R_0}$ and $i(B_r)|_{R_s}$ are the mean square radial components of the magnetic field on the photosphere and on the source surface, respectively.
We use in our calculations the classical approach assuming that the photosphere surface is potential and the source surface lies at 2.5 solar radii from the center. This approach was first proposed and justified in (\cite{H84, H91}). The use of a different scheme will somewhat change the magnetic-field values, but not the general relation of the harmonics. Thus, $\zeta = 0.4$.

\begin{figure}
\includegraphics[width=0.30\columnwidth]{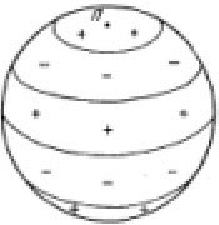}
\includegraphics[width=0.33\columnwidth]{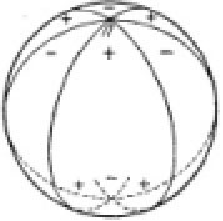}
\includegraphics[width=0.33\columnwidth]{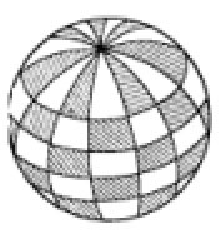}
\caption{Examples of different structures (from left to right) -- zonal, sectorial, and tesseral structures.} 
\label{f2}
\end{figure}

Based on the harmonic coefficients, one can isolate the following large-scale magnetic field structures \citep{OY89, Setal89, OS92} (Fig.~\ref{f2})  (here we give formal description of the structures and explain its physical meaning in the next para):

\begin{itemize}
\item Zonal structures $(m=0)$, including the structures with the dipole symmetry (odd $l$, $l=1$ is the axial dipole) marked ZO and with the quadrupole symmetry (even $l$, $l=2$ is the axial quadrupole) marked ZE.
\item Sectorial structures $(m=1)$, including SO  (odd $l$, $l=m=1$ for the equatorial dipole and bisectoral structure) and SE (even $l$, the simplest case $l=m=2$).
\item Tesseral structures with $l \ne m$, which were practically not considered in the previously published literature.  
\end{itemize}

The physical meaning of the partial field indices is as follows: the zonal odd index ZO denotes a part of the field with a zonal odd type of symmetry. These harmonics are sometimes called dipole-like, since the most powerful of them is a dipole oriented along the axis of rotation. The zonal parity index ZE determines the asymmetry of the magnetic field and is usually small due to the Hale`s effect. The sector indices SO and SE are usually associated with active longitudes on the Sun and the sector structure of the interplanetary magnetic field.

The zonal and sectorial harmonics are the structures with a spatial scale in one direction (in latitude or in longitude) comparable with the size of the whole Sun. Obviously, they can not be directly associated with local magnetic structures, though they are somehow related to them in the generation of the magnetic field. On the contrary, the tesseral modes have a limited spatial size and can be, in principle, associated with local magnetic fields. Of course, the quality of observational data does not allow us to say this with certainty,  because the low resolution of the source data (2 arcmin) does not allow the higher-order harmonics {\it l} and m to be calculated accurately enough.

\section{Relating magnetic fluxes and energies with particular harmonics}

The advantage of the above indices was demonstrated by \cite{OS92},  who applied them to the analysis of Stanford data with 3 arc min resolution for the period from June 1976 to September 1985, i.e. for Carrington rotations No. 1641-1766. The indices proved to be relevant for characterizing the solar cycle reference points. Modern data are available for 1976-2022, i.e. for 4 cycles. We have carried out the corresponding analysis by calculating the root mean square value of $i(B_r)$ for each half of the Carrington rotation and, then, performed a smoothing over 27 points, i.e., a year (Fig.~3).  

\begin{figure}
\includegraphics[width=0.97\columnwidth]{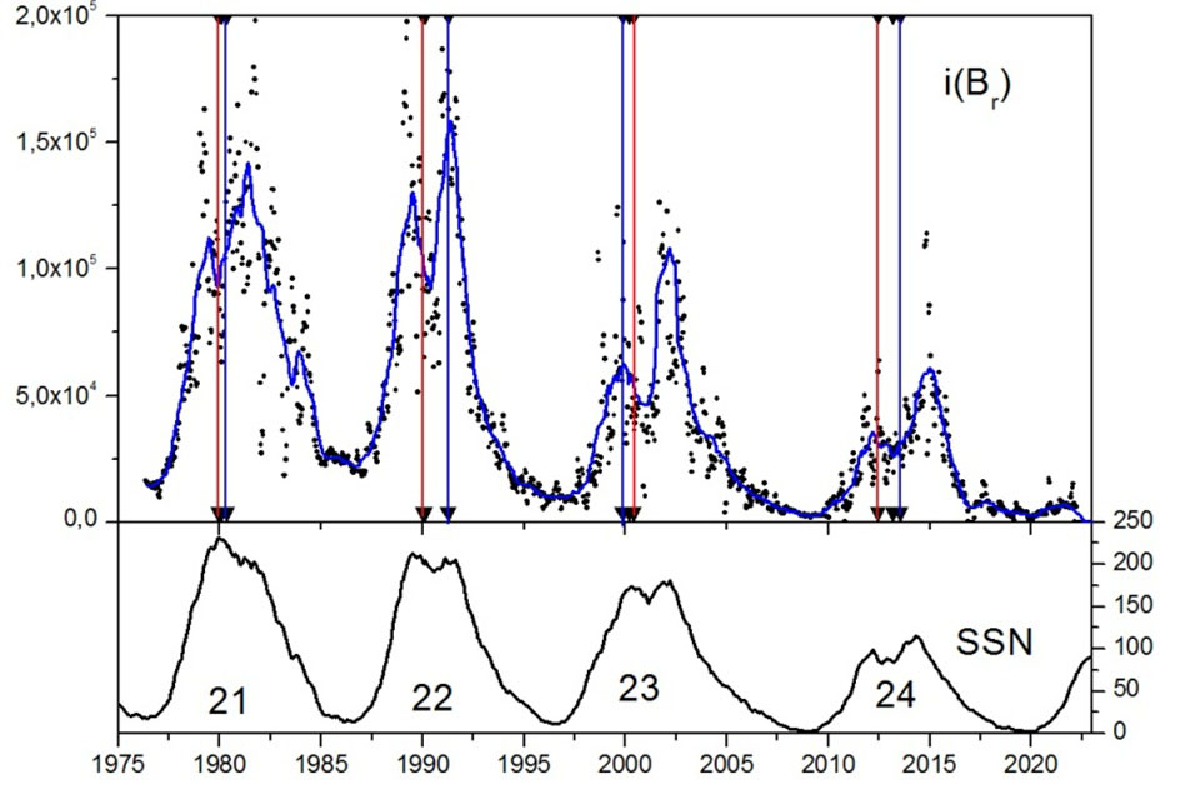}
\caption{Evolution of $i(B_r)$ at the photoshere level during the time interval 1976--2022 (upper panel in $\mu T^2$) compared to sunspot data (lower panel, cycle numbers are indicated, SSN are given on the right). The vertical red line shows the instant of magnetic field reversal at the North pole, the blue line stands for the reversal at the South pole. } 
\label{f3}
\end{figure}

As seen from Fig.~\ref{f3}, the time variation of $i(B_r)$ is quite similar to that of sunspot numbers. The cycle amplitudes gradually decay with time, but the shapes of the cycles on the two panels are somewhat different. Both cycles display a double structure; however, the second maximum in sunspots is well pronounced only in Cycle 24, while the second maximum in $i(B_r)$ is clearly seen in both cycles and is higher than the first one. The Gnevyshev gap in $i(B_r)$ exactly coincides with the reversals of the polar magnetic field. 

Below, we discuss the $i(B_r)$ index and its various modifications on the photosphere and at the source surface. The point is that the index on the photoshere includes the local magnetic fields while the index at the source surface represents mainly the large-scale field, i.e., the harmonics with lower $l$. Formally, this follows from Eqs.~(\ref{iph},\ref{is}), where the dependence on $\zeta$ is quite strong. Physically, this means that the large-scale magnetic fields decay with height much slower than the small-scale fields. This is why we consider the magnetic field structures on the source surface to be global. The index can be used, in particular, to identify the solar cycle reference points (e.g., see \cite{OS92} and references therein  where the local magnetic structures are mainly discussed). Here, we extend the discussion to include  the global field indices as well (see Figs. 4 and 5).

Our conclusions concerning the cycle reference points can be summarized as follows.

When describing the solar cyclic variation, it is customary to designate the main phases of the solar cycle as $m$ (the main phase), $A$ (the ascending branch) , $M$ (the maximum phase), and $D$ (the descending branch). Each of these phases lasts several years and their precise timing is difficult. It seems more convenient to introduce the concept of reference points separating these phases:

\begin{itemize}
    \item 
$t_{mA}$ - the end of the minimum phase and the beginning of the ascending branch,

\item
$t_{AM}$ - the beginning of the maximum phase

\item
$t_{MD}$ - the end of the maximum phase, and the beginning of the descending branch

\item
$t_{Dm}$ - the beginning of the minimum phase.
\end{itemize}

The concept of these reference points was first proposed by \cite{Vetal86} and, then, developed  by \cite{OS92}, who demonstrated that the reference points are the periods of sharp changes in all solar activity indices.

The reference point  $t_{mA}$ corresponds to the ascending branch of the cycle, it lasts 1-2 years after the minimum of the sunspot cycle. During this period, the spots of the previous cycle finally disappear, the spot formation zone shifts toward  the equator, and the global energy of the magnetic field $i(B_r)$ increases sharply.  Simultaneously, all indices on the photosphere grow, which indicates a substantial contribution from the local magnetic field. The indices at the source surface demonstrate a quite different behaviour. The general decay of $i (B_r)_{SS}$ is associated with an abrupt decrease in the contribution of the zonal and, especially, the odd (dipole-like) harmonics to $ZO_{SS}$ (configuration at the source surface), while the contribution of the sectorial and tesseral harmonics of the global magnetic field grow. The latter looks quite unexpected.

The reference point $t_{AM}$ takes place about a year before the cycle maximum and coincides with the start of the reversal of the polar magnetic field (see for comparison Figs.~\ref{f1}a and 3). This is the time of a local maximum in the number of active regions and the total number of flares. This point was called  "pre-maximum". At the same time, the total energy of the global field reaches a local minimum. The $ZO_{SS}$ field, which was predominant at the beginning of the cycle declines to minimum.  Simultaneously, a sharp increase in the $SO_{SS}$ field begins, leading to a pronounced maximum of the SO field and to a less pronounced maximum of the SE field after several solar rotations. 

This point is interesting as a point where the harmonics with $l=5$  become maximum. (See Figs.~1, 2 in \cite{Oetal23}, where the the authors proposed to call this point ”the overlapping phase” and to use it for predicting the date of maximum of Cycle 25). Based on the position of this point in mid-2022, we expect the sunspot maximum in Cycle 25 with a height 125 at the end of
2023.  Note that \cite{P24} estimated the height as $130.7 \pm 0.5$.

\begin{figure}
\includegraphics[width=0.97\columnwidth]{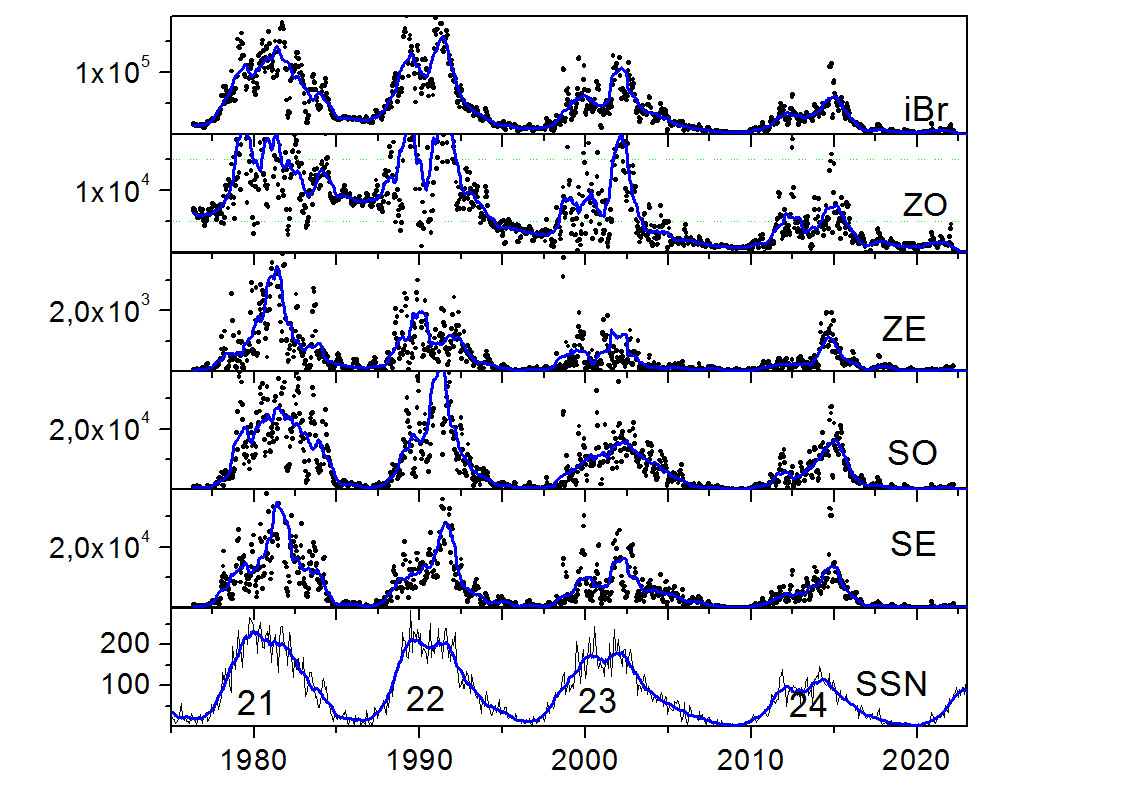} \\
\includegraphics[width=0.97\columnwidth]{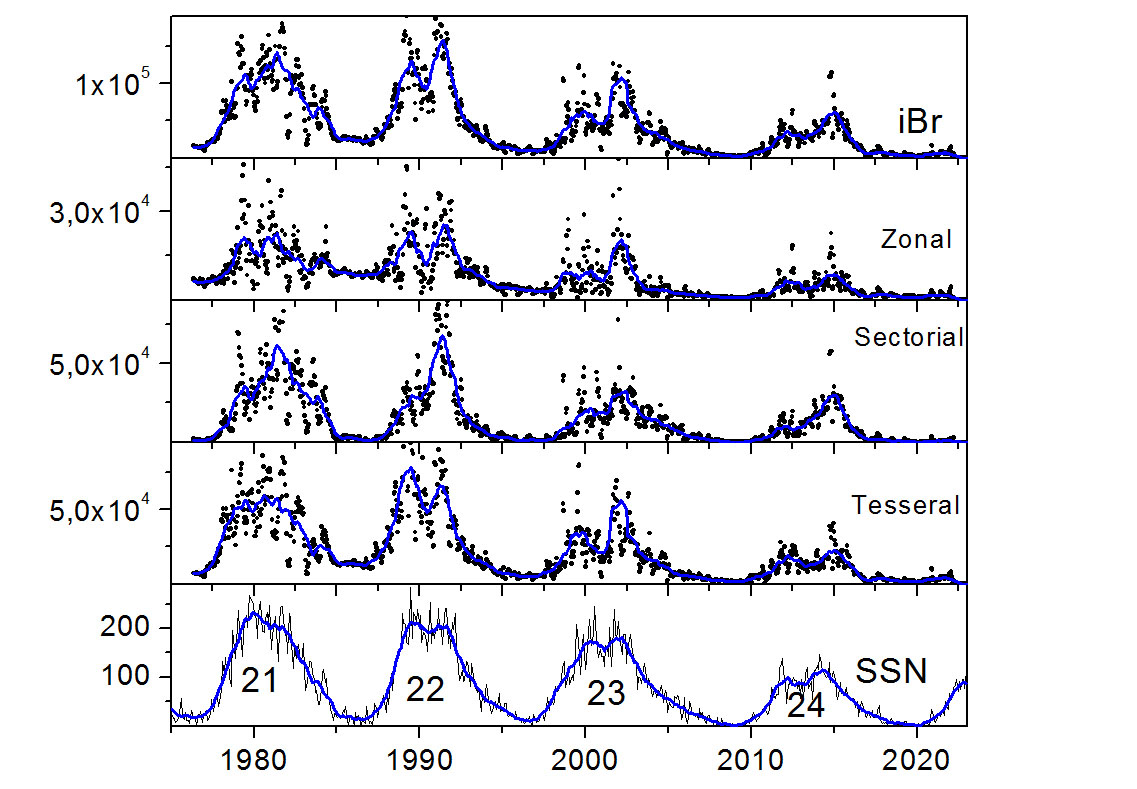}
\caption{Contribution of different harmonics to the $i(B_r)$ index as calculated at the photosphere level for 1976--2022. Upper panel: second row from top -- contribution of zonal odd harmonics,  third row from top -- sectorial odd harmonics,  fourth row from top -- sectorial even harmonics. Lower panel:  second row from top - contribution of zonal harmonics, third row from top - contribution of sectorial harmonics, forth row from top -- contribution of tesseral harmonics. For ease of comparison, the first and the last rows on both panels represent, respectively, the total  $i(B_r)$ index and $SSN$.} 
\label{f4}
\end{figure}

Fig.~\ref{f4} clearly shows that the second maximum in each cycle is determined by the contribution of the sectorial harmonics. The fact is that the second maximum occurs in the period of equatorial position of the global solar dipole and  formation of two-sector structure of the interplanetary magnetic field. This is the period when particularly large active regions appear on the Sun.

\begin{figure}
\includegraphics[width=0.97\columnwidth]{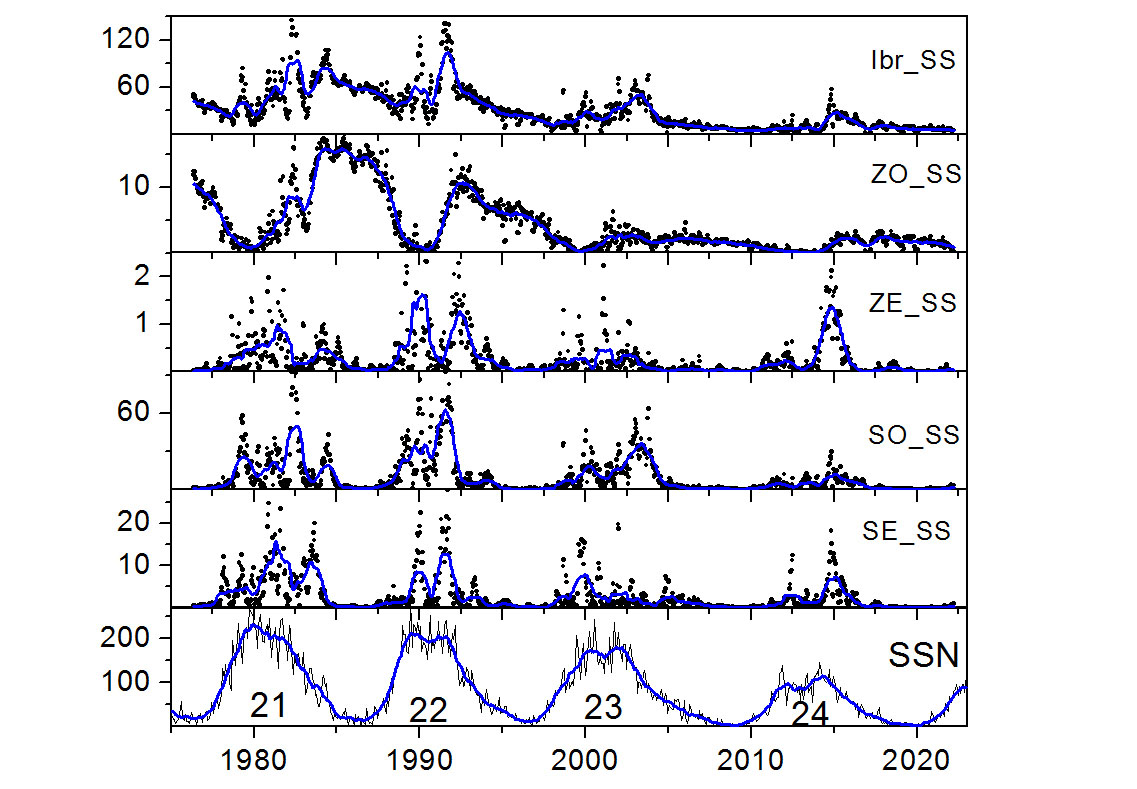} \\
\includegraphics[width=0.97\columnwidth]{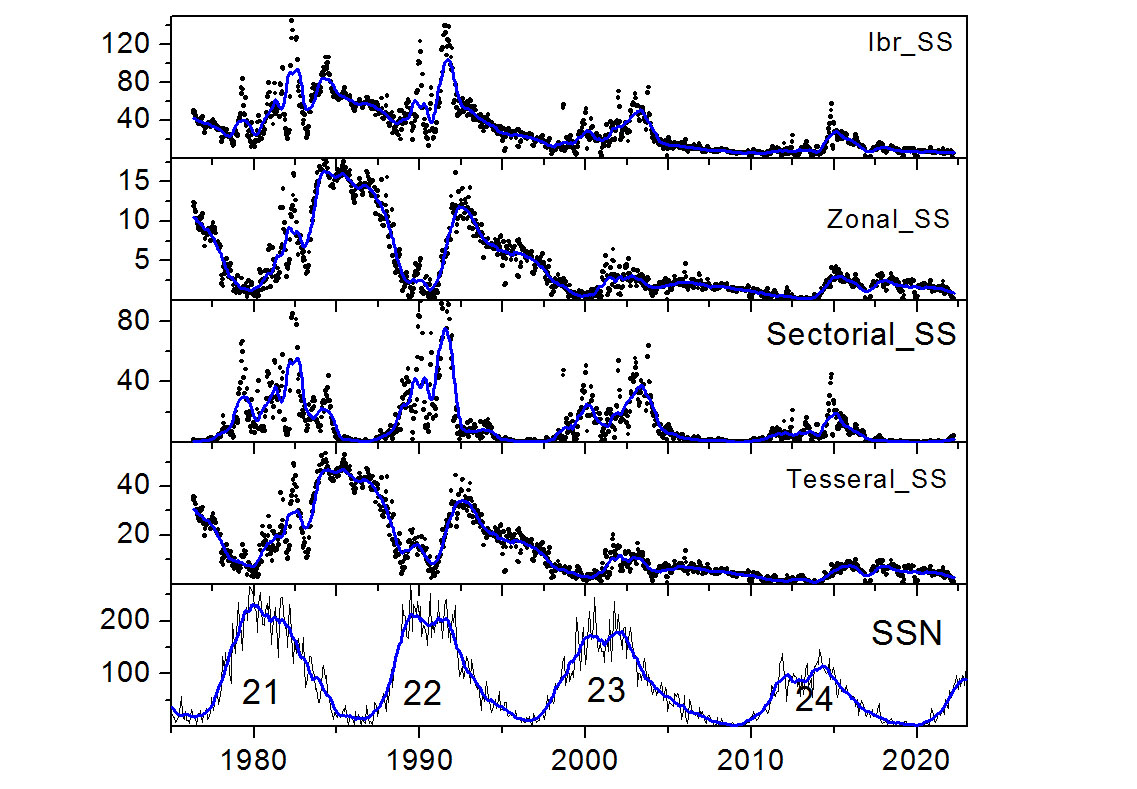}
\caption{Contribution of different harmonics to the $i(B_r)$ index as calculated at the source surface. All symbols are the same as in Fig.~\ref{f4}.} 
\label{f5}
\end{figure}

However, the second maximum is also pronounced in the contribution of the zonal odd (ZO) as well as in the sectorial even (SE) harmonics (Fig.~\ref{f4}). The latter is responsible for the 4-sector structure of the interplanetary magnetic field.
The {\bf maximum phase} (M) covers the whole interval of the sign reversal. In the maximum phase, all indices calculated at the photosphere level become maximum.  However, the  Wolf number maximum is determined basically by the behaviour of local magnetic fields; therefore, the variation of all indices on the source surface is more pronounced (Fig.~\ref{f5}).

Another feature seen on the plot is that the global magnetic field index $ZO_{SS}$ decreases, and the field minimum closely coincides with the date of maximum of the Wolf numbers. At the beginning of the {\bf descending phase},  $t_{MD}$, the ZO at the source surface is gradually restored, which looks like the generation of a new global magnetic field. In terms of the global magnetic field, the descending phase takes about ten rotation and contains some fine structure inside. { \bf The beginning of the minimum phase}  $t_{Dm}$ is characterized by the appearance of the first high-latitude sunspot groups belonging to a new cycle,  accompanied by a sharp decrease in the number of active regions of the old cycle.

To summarize, the global field indices allow us to locate the reference points of the cycle. These points were introduced earlier based on general characteristics of the solar cycle (\cite{Vetal86}),  but they were considered difficult to locate on the Wolf number curve.  After the points have been identified with the aid of the global magnetic-field indices, one can see that the Wolf number curve  smoothed over six solar rotations displays some features that coincide with the reference points of the cycle. 

As to the cycle amplitudes, all indices calculated at the photosphere surface behave more or less like the Wolf numbers. The general trend in Figs.~\ref{f3} - \ref{f5} that is most interesting in the context of prediction of the solar cycle amplitude is its gradual decay. There are, however, some remarks to be made. 

According to the basic polarity law, the northern and southern hemispheres are antisymmetric in sign. This means that the asymmetry must be very small and is revealed using local-field data. It is clear from Figs. 4 and 5  that ZE is always much smaller than ZO, which means that the polarity law is global and is true  both on relatively small and on large scales, both at the photosphere level and at the source surface.

The second remark concerns the well-known Gnevyshev-Ohl rule, according to which the even-numbered cycles are always lower than the following odd-numbered cycles \cite{GO48}. This rule is probably not entirely correct when applied to the global magnetic field. In any case, it was broken in the couple of Cycles 22-23. It was shown in a number of publications, e.g. \cite{D03, H10, KO14, ZP15, NO18}. 

\section{Relative contribution of harmonics of different symmetries to the magnetic energy}

\begin{figure}
\includegraphics[width=0.97\columnwidth]{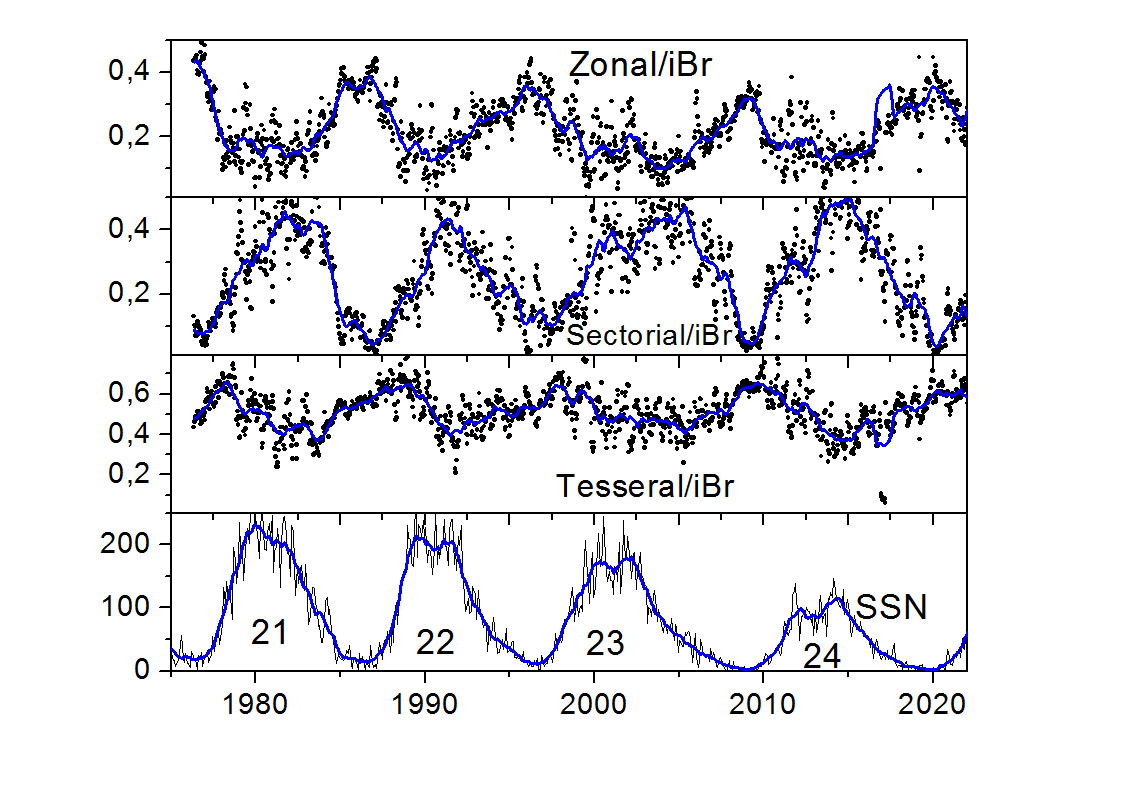} 
\caption{Contribution of different harmonics to the magnetic energy as calculated  at the photosphere level. The symbols are the same as in Fig.~\ref{f4}.} 
\label{f6}
\end{figure}

\begin{figure}
\includegraphics[width=0.97\columnwidth]{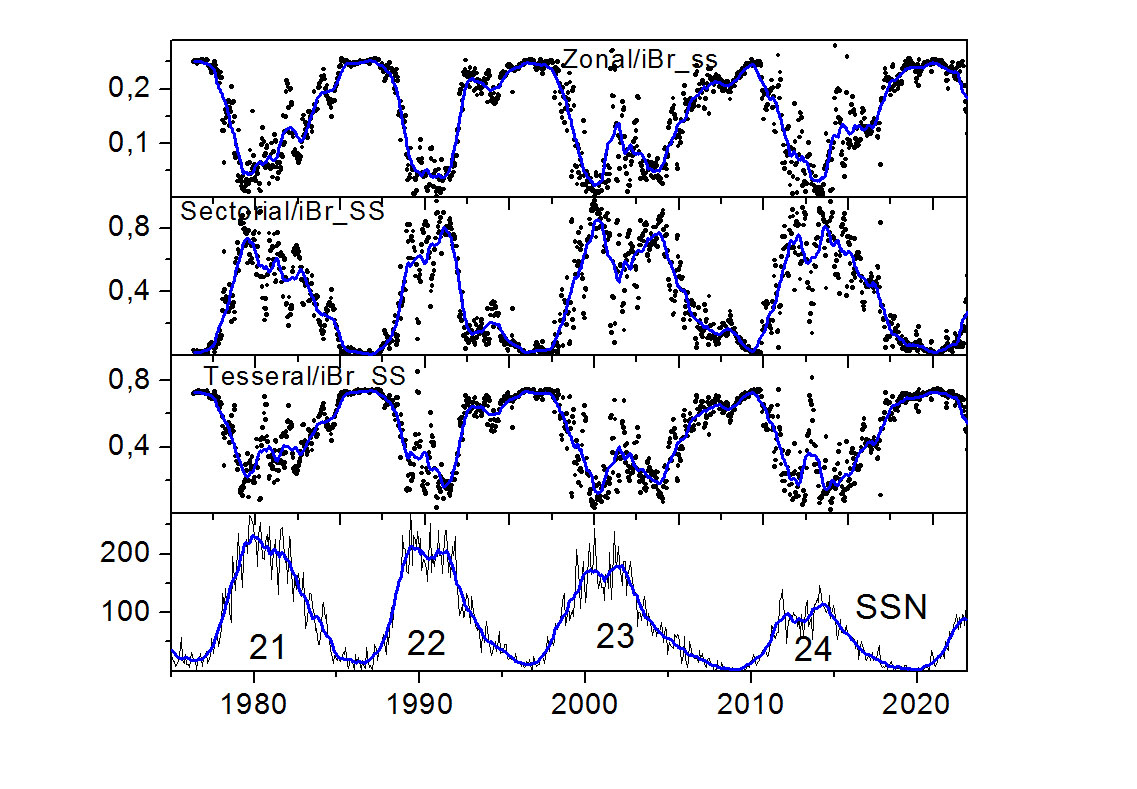} 
\caption{Contribution of different harmonics to the magnetic energy as calculated at the source surface. The symbols are the same as in  Fig.~\ref{f4}.} 
\label{f7}
\end{figure}

Apart from the prediction purposes, it is interesting to compare the contribution of different types of harmonics to the magnetic energy. In Fig. 3-5, one can see a gradual decrease in the number of sunspots, as well as in the total energy of the magnetic field and all its components during four cycles. Therefore we use a different normalization procedure relating the energy contributed by each harmonic to the total magnetic field energy (see Fig.~\ref{f6} for the indices on the photosphere  including the contribution of local fields and Fig.~\ref{f7} for the source surface, i.e. for the magnetic fields of larger spatial scales). 

It is quite surprising to see that, after normalization, the trend from Cycle 22 to Cycle 24 is absent, and the curves of all energy indices are very similar both in amplitude and in shape. 

It is noteworthy that the contribution of the zonal and sectorial modes to the magnetic energy occurs in antiphase. On the contrary, the contribution of the tesseral harmonics remains almost not modulated and only slightly increases in the epochs of the cycle minimum. All these factors result in violation of the Gnevyshev-Ohl rule, which, apparently, implies that the rule is associated with the secular modulation of the SNN cycle. 

To summarize, we can say that each cycle has its own hierarchy of structures with different types of symmetry. This is, apparently, due to the fact that the properties of the solar dynamo do not depend on the total magnetic energy in a given cycle. The trend in cycle amplitudes is obviously determined by a different mechanism.

\section{Conclusion: Magnetic scenario of a solar cycle}

Based on the above results, we can suggest a universal quantitative scenario of a solar cycle, which does not depend on the height of the cycle or on its number.

At the beginning of the cycle, the zonal harmonics on the potosphere are responsible for approximately 37- 42\% of the total magnetic energy  (cf. with 100\% in oversimplified models). The sectorial modes remain significant and account for about  5-10\% of the general magnetic energy. The main contribution to the magnetic energy (about 40\%) is made by tesseral harmonics, which are not directly related to either the number of sunspots or the global magnetic field.  As the cycle develops, the relative contribution of the zonal harmonics gradually decreases and becomes minimum (about 15-18\%) immediately before the SSN maximum.  On the contrary, the relative contribution of the sectorial harmonics  increases and becomes maximum (60-65\%) slightly after the sunspot maximum, at the time of the secondary SSN maximum or even at the beginning of the decay phase. As for the tesseral harmonics, their contribution becomes maximum (60\% -- 65\%) just before the maximum in sunspot numbers. After that, it decreases slightly to reach its typical value of about 40\% and, then, grows again. 

This scenario is  illustrated in Fig.~7 for contributions to the magnetic energy at the source surface, which presents the  contribution of the large-scale magnetic field (of course, with slightly different figures). Note that the contribution of  tesseral harmonics to the total magnetic energy in the epochs of SSN minimum turns out to be surprisingly high and stable (about 70\%). 

\section{Discussion}

Above, we have examined the cyclic variations in the magnetic field and magnetic energy during the past four cycles of solar activity (Cycles 21-24). Unlike the previous work \cite{Oetal23}, where the contribution of zonal harmonics of the large-scale  magnetic field was considered, this study includes all types of symmetries: zonal, sectorial and tesseral. An important part of the analysis is devoted to the energy indices proposed by \cite{OY89, Setal89, OS92}, cf. \cite{Zetal19}. The indices represent the energy of each magnetic field component (more precisely, the corresponding mean square value of the radial magnetic field) and can be calculated at any distance from the solar surface using the data at the photosphere level. The use of energy indices turns out to be more informative than a direct analysis of the magnetic-field intensity data.

From the above analysis of observational data, we can draw the following conclusions regarding the solar dynamo.

Strictly speaking, the solar dynamo generates a uniform cycle with a quite complicated internal structure. Different components of the structure make their particular stable contribution to the magnetic energy measured in terms of $i(B_r)$. We do not see any signs of the dynamo as a mechanism responsible for different heights of the cycles. One has to include something apart from the classical dynamo drivers in the scenario to explain the appearance of solar cycles with different heights. The simplest way would be to include stochastic oscillations of the dynamo drivers. The point, however, is that the role of purely statistical fluctuations obviously decreases with the increasing number of the compared cycles and can hardly explain long-term modulation of a cycle, similar to the Gleisberg cycle. 

The phenomenon of cyclic solar activity is supposed to include tree different, yet, connected processes:

1. A classic 11-year cycle controlled by differential rotation, when the poloidal magnetic field is transformed into the toroidal one, and a certain mirror-asymmetric mechanism that restores the poloidal magnetic field from the toroidal one. Taken independently, this process produces cycles that are similar to each other.

2. A specific process producing sunspots from the magnetic field generated by solar dynamo. In the framework of the flux-transport dynamo the process produces mirror asymmetry crucial for the dynamo action however it do not determine how this mirror asymmetry plays its role. 

3. Some kind of external mechanism responsible for a long-term cycle modulation. This mechanism is postulated, say, by \cite{NP20}.

The question is how well the scenario based on the analysis of the last few cycles is consistent with the solar dynamo theory and the existing  knowledge of the previous solar activity cycles. In mathematical terms, the scenario means that the total solar magnetic field $\bf B$ can be divided into several components:

\begin{equation}
{\bf H} =  Q {\bf B} + {\bf b}.   
\end{equation}
We follow the ideas of \cite{NP20} however we are constraining its possible form basing on observational data rather suggest a particular model equation suitable to mimic long-term solar activity modulation.

Here, $\bf B$ represents the part of the total magnetic field involved in the 11-year cycle. First of all, it includes zonal-odd harmonics with the leading role of the harmonic with $l=5$ followed by other modes. The standard separation of the governing equations for the magnetic field $\bf B$ from the general equations of solar MHD was proposed by \cite{P55} and is included in formalization of all other solar dynamo models. Of course, the physical interpretation of the governing parameter in a model is specific to each model. 

For nonlinear oscillations, the separation into the part responsible for oscillation (here, $\bf B$) and the envelope $Q$ with a slow evolution is a completely standard procedure used in many branches of physics, e.g., in nonlinear optics \citep{B96}. Available analytical experiences with nonlinear migratory dynamos (e.g., \cite{M97}) support this expectation.

The point however is that  the long-term evolution of the solar cycle is a complex phenomenon, which involves a lot of timescales, e.g.,the Gleissberg cycle of about a hundred years (\cite{U17}). In order to obtain this long-term dynamics, the equation for the envelope $Q$ has to include some contribution from an external physical factor. The choice of such a phenomena is not too easy. There is a proposal to connect the dynamics with tidal perturbations from Jupiter, which requires an orbital period close to the length of a cycle. Indeed, the orbital period of Jupiter (11.682 yrs) is close to the nominal length of a solar cycle. Due to the beating effect, this  allows us to expect timescales of 100-200 yrs and 5--6 yrs (see, e.g., \cite{Setal23}). The corresponding tidal forces are, however, rather weak. The specific way to support the desired dynamics remains unclear, and the search for promising examples of this kind in stellar cycle observations leads to negative results (\cite{Oetal22}). Numerical simulation  \citep{OFS23} confirm that statistical fluctuations in the dynamo drivers easily allow us to get more or less realistic cycle variations over the timescales of 5-6 years; however, it is problematic to get the scales of 100 -- 200 yrs.  Dynamical chaos in the amplitude equations looks as a realistic way to fit simultaneously short- and long-term dynamics of the solar cycle amplitude.
Of course, involving leptocline as a possible domain for dynamo action in addition to tachocline opens here more options to explain complicated structure of solar activity cycle.

\section*{Acknowledgments}
VNO and DDS acknowledge the support of the Ministry of Science and Higher Education of the Russian Federation under the grant 075-15-2020-780 (VNO) and 075-15-2022-284 (DDS). DDS  and AAS thanks support by BASIS fund number 21-1-1-4-1. 

\section*{Data availability statement.}
The main data series used in the analysis were received from the John Wilcox Stanford Observatory (WSO). 
These data began in May 1976 (Carrington Rotation 1641) and have been continued till the present (http://wso.stanford.edu/forms/prsyn.html). The polar field is given after http://wso.stanford.edu/Polar.html. 

The sunspot data were taken from https://www.sidc.be /silso/datafiles

\bibliographystyle{mnras}
\bibliography{sample631}

\end{document}